\documentclass{llncs} 

\usepackage{amssymb}
\usepackage{textcomp}
\usepackage{comment}
\usepackage{color}
\usepackage{graphicx}
\usepackage{amsmath}
\newcommand{\Beginproof}{{\em Proof.}  }
\newcommand{\Endproof}{\hfill$\Box$\\}

\newcommand{\knobdd}{$k$-{\bf NOBDD} }

\newcommand{\kpobdd}{$k$-{\bf POBDD} }

\newcommand{\kqobdds}{$k$-QOBDDs }

\newcommand{\ket}[1]{|#1\rangle}

\begin{document}

\title{Width Hierarchies for Quantum and Classical Ordered Binary Decision Diagrams with Repeated Test}
\author{Kamil~Khadiev$^{1,2,}$\thanks{Partially supported by ERC Advanced Grant MQC. The work is performed according to the Russian Government Program of Competitive Growth of Kazan Federal University}\and Rishat~Ibrahimov$^2$}

\institute{ University of Latvia, Riga, Latvia      
       \and       
Kazan Federal University, Kazan, Russia      
                \\ \email{kamilhadi@gmail.com, rishat.ibrahimov@yandex.ru} 
}

\maketitle

\begin{abstract}
We consider quantum, nondterministic and probabilistic versions of known computational model Ordered Read-$k$-times  Branching Programs or Ordered Binary Decision Diagrams with repeated test ($k$-QOBDD, $k$-NOBDD and $k$-POBDD). We show width hierarchy for complexity classes of Boolean function computed by these models and discuss relation between different variants of $k$-OBDD.

\end{abstract}

\section{Introduction}
Branching programs are one of the well known models of computation. These models have been shown useful in a variety of domains, such as hardware verification, model checking, and other CAD applications (see for example the book by I. Wegener \cite{Weg00}). It is known that the class of Boolean  functions computed by polynomial size branching programs coincide with the class of functions computed by non-uniform log-space machines.

One of important restrictive branching programs are oblivious read once branching programs, also known as Ordered Binary Decision Diagrams (OBDD) \cite{Weg00}. It is a good model of data streaming algorithms. These algorithms are actively used in industry, because of rapidly increasing of size of data which should be processed by programs. Since a length of an OBDD is at most linear (in the length of the input), the main complexity measure is ``width'', analog of size for automata.
And it can be seen as nonuniform automata (see for example \cite{ag05}).  

In the last decades quantum model of OBDD came into play \cite{agk01}, \cite{nhk00}, \cite{SS05}, \cite{s06}. Researchers also interested in read-$k$-times  quantum model of OBDD ($k$-QOBDD). $k$-QOBDD can be explored from automata point of view. And in that case we can found good algorithms for two way quantum classical automata in paper \cite{AW02} of Ambainis and Watrous. Other automata models, that have relation with $k$-QOBDD are restart and reset quantum automata  \cite{YS10B}.

One of the interesting questions, which researchers explored is hierarchy of complexity classes for classical and quantum $k$-OBDDs. These models have two main characteristics of complexity: number of layers ($k$) and width. Hierarchy for numbers of layers was investigated in papers \cite{bssw96},  \cite{K16}, \cite{kk2017}, \cite{akk2017}

In same time, there are only few work on width hierarchy. For example, width hierarchy for deterministic $k$-OBDD is presented in \cite{k15}. Width hierarchies for classical and quantum $1$-OBDDs  are discussed in \cite{agky14}, \cite{agky16}, \cite{kk2017}.

In this paper we prove width hierarchy for nondeterministic, probabilistic $k$-OBDD and quantum $k$-OBDD with natural order of input bits. We considered sub linear width and hierarchies based on results and lower bounds from \cite{AK13}, \cite{K16}, \cite{akk2017}.

The paper is organized in following way. In Section \ref{sec:prlmrs} we present definitions. Section \ref{sec:s1} contains Hierarchy results for classical models and Section \ref{sec:s2} for quantum one. We discuss relation between different models in Section \ref{sec:s3}.

\section{Preliminaries}\label{sec:prlmrs}

Ordered  read ones  Branching Programs (OBDD) are well known model for Boolean functions computation. A good source for different models of branching programs is the book by I. Wegener  \cite{Weg00}.

A branching program  over a set $X$ of $n$ Boolean variables is
a directed acyclic graph with two distinguished nodes $s$ (a source node) and $t$ (a sink node). We denote  such program $P_{s,t}$ or just $P$.   Each inner node $v$  of $P$ is associated with a variable $x\in X$. {\em Deterministic} $P$ has exactly two outgoing edges labeled $x=0$   and $x=1$ respectively for such node $v$.

The program $P$ computes the Boolean function $f(X)$ ($f:\{0,1\}^n \rightarrow \{0,1\}$) as follows: for each $\sigma\in\{0,1\}^n$ we let $f(\sigma)=1$ if and only if there exists at least one $s-t$ path (called {\em accepting} path for $\sigma$) such that all edges along this path are consistent with $\sigma$.

A branching program is {\em leveled} if the nodes can be partitioned into levels $V_1, \ldots, V_\ell$ and a level $V_{\ell+1}$ such that the nodes in $V_{\ell+1}$ are the sink nodes, nodes in each level $V_j$ with $j \le \ell$ have outgoing edges only to nodes in the next level $V_{j+1}$. For a leveled $P_{s,t}$ the source node $s$ is a node from the first level  $V_1$ of nodes  and the sink node $t$ is a node from the last level $V_{\ell+1}$.

The {\em width} $w(P)$ of a leveled branching program $P$ is the maximum
of number of nodes in  levels of $P$. $ w(P)=\max_{1\le j\le \ell}|V_j|. $ The {\em size} of branching program $P$ is a number of  nodes of
program $P$.

A leveled branching program is called {\em oblivious} if all inner
nodes of one level are labeled by the same variable.  A branching
program is called {\em read once} if each variable is tested on each
path only once. 
An oblivious leveled read once branching program is also called Ordered  Binary Decision Diagram (OBDD).
OBDD $P$ reads variables in its individual  order
$\pi=(j_1,\dots,j_n)$, $\pi(i)=j_i$, $\pi^{-1}(j)$ is position of $j$ in permutation $\pi$. We call $\pi(P)$ the order of $P$. Let us denote natural order as $id=(1,\dots,n)$. Sometimes we will use notation $id$-OBDD $P$, it means that $\pi(P)=id$. Let $width(f)=\min_{P}w(P)$ for OBDD $P$ which computes $f$ and $id\!-\!width(f)$ is the same but for $id$-OBDD.

The Branching program $P$ is called $k$-OBDD if it consists from $k$ layers, where  $i$-th ($1\le i\le k$) layer  $P^i$ of $P$  is  an OBDD. Let  $\pi_i$ be an order of $P^i$, $1\le i\le k$ and $\pi_1=\dots=\pi_k=\pi$.
 We call  order
$\pi(P)=\pi$ the order of $P$.

Nondeterministic OBDD (NOBDD) is nondeterministic counterpart of OBDD.
Probabilistic OBDD (POBDD) can have more than two edges for node, and choose one of them using probabilistic mechanism. POBDD $P$ computes Boolean function $f$ with bounded error $0.5-\varepsilon$ if probability of right answer is at least $0.5+\varepsilon$.


Let us discuss a definition of quantum OBDD (QOBDD). It is given in different terms, but you can see that it is equivalent. You can see \cite{agkmp2005}, \cite{agk01} for more details.

For a given $ n>0 $, a quantum OBDD $ P$ of width $w$, defined on $ \{0,1\}^n $, is a 4-tuple
$
	P=(T,\ket{\psi}_0,Accept,\pi),
$
where
\begin{itemize}
	\item $ T = \{ T_j : 1 \leq j \leq n \mbox{ and } T_j = (G_j^0,G_j^1)  \} $ are  ordered pairs of (left) unitary matrices representing the transitions is applied at the $j$-th step, where $ G_j^0 $ or $ G_j^1 $, determined by the corresponding input bit, is applied.
	\item $\ket{\psi}_0$ is initial vector from $ w $-dimensional Hilbert space over field of complex numbers.  $ \ket{\psi}_0=\ket{q_0}$ where $ q_0 $ corresponds to the initial node.
	\item $ Accept \subset \{1,\ldots,w\} $ is accepting nodes.
\item $ \pi $ is a permutation of $ \{1,\ldots,n\} $ defining the order of testing the input bits.
\end{itemize}

  For any given input $ \sigma \in  \{0,1\}^n $, the computation of $P$ on $\sigma$ can be traced by  a vector from $ w$-dimensional Hilbert space over field of complex numbers. The initial one is $ \ket{\psi}_0$. In each step $j$, $1 \leq j \leq n$, the input bit $ x_{\pi(j)} $ is tested and then the corresponding unitary operator is applied:
$
	\ket{\psi}_j = G_j^{x_{\pi(j)}} (\ket{\psi}_{j-1}),
$ 
where $ \ket{\psi}_{j-1} $ and $ \ket{\psi}_j $ represent the state of the system after the $ (j-1)$-th  and $ j$-th steps, respectively, where $ 1 \leq j \leq n $.

In the end of computation program $P$ measure qubits. The accepting (return $1$) probability $Pr_{accept}(\sigma)$ of $ P_n $ on input $ \sigma $ is 
$
	Pr_{accept}(\nu)=\sum_{i \in Accept} v^2_i.
$, for $ \ket{\psi}_n=(v_1,\dots,v_w)$.
We say that a function $f$ is computed by $ P$ with bounded error if there exists an $ \varepsilon \in (0,\frac{1}{2}] $ such that $ P$ accepts all inputs from $ f^{-1}(1) $ with a probability at least $ \frac{1}{2}+\varepsilon $ and $ P_n $ accepts all inputs from $ f^{-1}(0) $ with a probability at most $ \frac{1}{2}-\varepsilon $.  We can say that error of answer is $\frac{1}{2}-\varepsilon$.

Let $k$-{\bf QOBDD}$_{w}$ be a set of Boolean functions which can be computed by bounded error $k$-QOBDDs of width $w$. $k$-id-{\bf QOBDD}$_{w}$ is same for bounded error $k$-QOBDDs with order $id=(1,\dots,n)$. $k$-{\bf NOBDD}$_{w}$ and $k$-{\bf POBDD}$_{w}$ is similar classes for $k$-NOBDD and bounded error $k$-POBDD.

\section{Width Hierarchies on Classical $k$-OBDD}\label{sec:s1}
Firstly, let us discuss required definitions.

 Let $\pi=(X_A,X_B)$
 be a partition of the set $X$ into two parts $X_A$ and $X_B=X\backslash X_A$. Below we will use equivalent notations $f(X)$ and $f(X_A, X_B)$.
Let  $f|_\rho$ be  a subfunction of $f$, where  $\rho$ is mapping $\rho:X_A \to \{0,1\}^{|X_A|}$.
Function $f|_\rho$ is obtained from $f$ by applying $\rho$. Let $N^\pi(f)$ be number of different subfunctions with respect to partition $\pi$.
Let $\Theta(n)$ be the set of all permutations of $\{1,\dots,n\}$.
Let partition $\pi(\theta,u)=(X_A,X_B)=(\{x_{j_{1}},\dots,x_{j_u}\},\{x_{j_{u+1}},\dots,x_{j_n}\})$, for permutation $\theta=(j_1,\dots, j_n)\in \Theta(n), 1<u<n$. We denote $\Pi(\theta)=\{\pi(\theta,u): 1<u<n\}$.
Let $ N^\theta(f)=   \max_{\pi\in \Pi(\theta)} N^\pi(f),
N(f)=\min_{\theta\in \Theta(n)}N^\theta(f). $ 

Secondly, let us present existing lower bounds for nondeterministic and probabilistic $k$-OBDDs. 

\begin{lemma}[\cite{AK13}]\label{lm:lower-kobdd}
Let function $f(X)$ is computed  by  $k$-OBDD $P$ of  width $w$,
then $N(f) \leq w^{(k-1)w+1}. $
\end{lemma}

\begin{lemma}[\cite{K16}]\label{lm:lower-knobdd}
Let function $f(X)$ is computed  by  $k$-NOBDD $P$ of  width $w$,
then $N(f) \leq 2^{w\big((k-1)w+1\big)}. $
\end{lemma}

\begin{lemma}[\cite{K16}]\label{lm:lower-pnobdd}
Let function $f(X)$ be computed  by bounded error  $k$-POBDD $P$ of  width $w$,
then 
\[N(f)\leq   \left(C_1k(C_2 +\log_2{w} + \log_2{k})\right)^{(k+1)w^2}\]
for some constants $C_1$ and $C_2$.
\end{lemma}

Thirdly, let us define {\em Shuffled Address Function} ($SAF_{k,w}$) from \cite{k15}  based on definition of well known Pointer Jumping Function \cite{bssw96}, \cite{nw91}.

\begin{definition}[Shuffled Address Function]
Let us  define Boolean function $SAF_{k,w}(X):\{0,1\}^n\to \{0,1\}$ for integer $k=k(n)$ and  $w=w(n)$ such that
\begin{equation}
 2kw(2w + \lceil \log k \rceil + \lceil \log 2w \rceil)<n.\label{kw}
\end{equation}
We divide input variables to $2kw$ blocks. There are $\lceil
n/(2kw)\rceil =a$ variables in each block.  After that we divide
each block to
 {\em address} and {\em value} variables. First  $\lceil\log k\rceil + \lceil\log 2w\rceil$ variables of block are {\em address}
 and other $a-\lceil\log k\rceil + \lceil\log 2w\rceil=b$ variables of block are {\em value}.

We call $x^{p}_{0},\dots,x^{p}_{b-1}$ {\em value} variables of $p$-th block and  $y^{p}_{0},\dots,y^{p}_{\lceil\log k\rceil + \lceil\log 2w\rceil}$ are {\em address} variables, for $p\in\{0,\dots,2kw-1\}$.

Boolean function $SAF_{k,w}(X)$ is iterative  process  based on definition of following six functions:

Function $AdrK:\{0,1\}^n\times\{0,\dots,2kw-1\}\to \{0,\dots,k-1\}$
obtains firsts part of block's address. This block will be used only
in step of iteration which number is computed using this function:

\[AdrK(X,p)=\sum_{j=0}^{\lceil\log k\rceil-1}y^{p}_{j}\cdot 2^{j} (mod\textrm{
}k).\]

Function $AdrW:\{0,1\}^n\times\{0,\dots,2kw-1\}\to \{0,\dots,2w-1\}$ obtains second part of block's address. It is the address of block within one step of iteration:

\[AdrW(X,p)=\sum_{j=0}^{\lceil\log 2w\rceil-1}y^{p}_{j+\lceil\log k\rceil}\cdot 2^{j} (mod\textrm{
}2w).\]

Function $Ind:\{0,1\}^n\times\{0,\dots,2w-1\}\times\{0,\dots,k-1\}\to \{0,\dots,2kw-1\}$ obtains number of block by number of step and address within this step of iteration:

\begin{displaymath}
Ind(X,i,t) = \left\{ \begin{array}{ll}
p, & \textrm{where $p$ is minimal number of block such that}\\
& \textrm{$AdrK(X,p)=t$ and $AdrW(X,p)=i$}, \\
-1, & \textrm{if there are no such $p$}.
\end{array} \right.
\end{displaymath}

Function $Val:\{0,1\}^n\times\{0,\dots,2w-1\}\times\{1,\dots,k\}\to \{-1,\dots,w-1\}$ obtains value of block which have address $i$ within $t$-th step of iteration:

\begin{displaymath}
Val(X,i,t) = \left\{ \begin{array}{ll}
\sum_{j=0}^{b-1}x^{p}_{j} (mod\textrm{ }w), & \textrm{where }p=Ind(X,i,t)\textrm{, for $p\geq 0$}, \\
-1, & \textrm{if }Ind(X,i,t)<0.
\end{array} \right.
\end{displaymath}

Two functions $Step_1$ and $Step_2$ obtain value of $t$-th step of iteration. Function $Step_1:\{0,1\}^n\times\{0,\dots,k-1\}\to \{-1,w\dots,2w-1\}$ obtains base for value of step of iteration:

\begin{displaymath}
Step_1(X,t) = \left\{ \begin{array}{ll}
-1, & \textrm{if }  Step_2(X,t-1)=-1, \\
0, & \textrm{if }  t=-1,\\
Val(X,Step_2(X,t-1),t) + w, & \textrm{otherwise}.
\end{array} \right.
\end{displaymath}

Function $Step_2:\{0,1\}^n\times\{0,\dots,k-1\}\to \{-1,\dots,w-1\}$ obtain value of $t$-th step of iteration:

\begin{displaymath}
Step_2(X,t) = \left\{ \begin{array}{ll}
-1, & \textrm{if }  Step_1(X,t)=-1, \\
0, & \textrm{if }  t=-1\\
Val(X,Step_1(X,t),t), & \textrm{otherwise}.
\end{array} \right.
\end{displaymath}

Note that address of current block is computed on previous step.

Result of Boolean function $SAF_{k,w}(X)$ is computed by following way:

\begin{displaymath}
SAF_{k,w}(X) = \left\{ \begin{array}{ll}
0, & \textrm{if }  Step_2(X,k-1)\leq 0, \\
1, & \textrm{otherwise}.
\end{array} \right.
\end{displaymath}
\end{definition}

Let us discuss complexity properties of the function:

\begin{lemma}[\cite{k15}]\label{lm:saf-lower}
For integer $k=k(n)$, $w=w(n)$ and Boolean function $SAF_{k,w}$,
such that inequality (\ref{kw}) holds, the following statement is
right: $N(SAF_{k,w})\geq w^{(k-1)(w-2)}$.
\end{lemma}

\begin{lemma}[\cite{k15}]\label{lm:saf-upper}
There is $2k$-OBDD $P$ of width $3w+1$ which computes $SAF_{k,w}$
\end{lemma}

Let us present Lemma \ref{lm:saf-lower} in more useful form:

\begin{corollary}\label{cr:saf-lower}
For integer $k=k(n)$, $w=w(n)$ and Boolean function $SAF_{k,w}$,
such that inequality (\ref{kw}) holds, the following statement is
right: $N(SAF_{k,w})\geq w^{kw/6}$.
\end{corollary}

\subsection{Hierarchy Results for Classical Models}

Hierarchy for deterministic OBDD is already known:

\begin{theorem}[\cite{k15}]\label{h-kobdd}
For integer $k=k(n),w=w(n)$  such that $2kw(2w + \lceil \log k
\rceil + \lceil \log 2w \rceil)<n, k\geq 2, w\geq 64$ we have
$k$-{\bf OBDD}$_{\lfloor w/16
\rfloor-3}\subsetneq k$-{\bf OBDD}$_{w}$.
\end{theorem}

Let us discuss hierarchies for nondeterministic and probabilistic models.

\begin{theorem}\label{th:hi-n}
For $w \geq 8$ we have $\knobdd _{\sqrt{w} / 2} \subsetneq \knobdd _{3w + 1}$
\end{theorem}
\Beginproof
It is clear that $\knobdd _{\sqrt{w} / 2} \subseteq \knobdd _{3w + 1}$. Let us proof inequality of these classes.
By Lemma \ref{lm:saf-upper} we have $SAF_{k,w} \in 2k$-{\bf NOBDD}$_{3w + 1}$. Let us show that $SAF_{k,w} \notin 2k$-{\bf NOBDD}$_{\sqrt{w} / 2}$. 

\[
	\frac{N(SAF_{k, w})}{2^{\sqrt{w}/2(1 + (2k - 1)\sqrt{w}/2)}} \geq
\]
\[
	\geq \frac{w^{kw/6}}{2^{\sqrt{w}/2(1 + (2k - 1)\sqrt{w}/2)}}=
\]

\[
= 2^{\frac{kw}{6}\log w - \frac{\sqrt{w}}{2} - \frac{1}{4}w(2k-1)} = 
\]

\[
= 2^{\frac{\sqrt{w}}{2}(\frac{k\sqrt{w}}{3}\log w - 1 - k\sqrt{w} + \frac{\sqrt{w}}{2})} = 
\]

\[
= 2^{\frac{\sqrt{w}}{2}(\frac{k\sqrt{w}}{3}(\log w - 3) + \frac{\sqrt{w}}{2} - 1)} > 1
\]

Therefore $SAF_{k,w} \not\in 2k$-OBDD$_{\sqrt{w} / 2}$, due to Lemma \ref{lm:lower-knobdd}. 

And $\knobdd _{\sqrt{w} / 2} \neq \knobdd _{3w + 1}$ 
\Endproof

\begin{theorem}\label{th:hi-p}
For $\sqrt{w}/ (\log_2 k \log_2w) \geq 1$ we have $\kpobdd _{\sqrt{w}/ (\log_2 k \log_2w)} \subsetneq \kpobdd _{3w + 1}$
\end{theorem}
\Beginproof
It is clear that $\kpobdd _{\sqrt{w}/ (\log_2 k \log_2w)} \subseteq \kpobdd _{3w + 1}$. Let us proof inequality of these classes.
By Lemma \ref{lm:saf-upper} we have $SAF_{k,w} \in 2k$-{\bf POBDD}$_{3w + 1}$. Let us show that $SAF_{k,w} \notin 2k$-{\bf POBDD}$_{\sqrt{w}/ (\log_2 k \log_2w)}$. 

\[
	\frac{N(SAF_{k, w})}{\left(C_1k(C_2 +0.5\log_2{w} - \log_2\log_2 k - \log_2\log_2w  + \log_2{k})\right)^{(2k+1)w/(\log_2 k \log_2w)^2}} \geq
\]
\[
	\geq \frac{w^{kw/6}}{\left(C_1k(C_2 +0.5\log_2{w} - \log_2\log_2 k - \log_2\log_2w  + \log_2{k})\right)^{(2k+1)w/(\log_2 k \log_2w)^2}} =
\]

\[
= 2^{\frac{kw}{6}\log w - ((2k+1)w/(\log_2 k \log_2w)^2)\log \left(C_1k(C_2 +0.5\log_2{w} - \log_2\log_2 k - \log_2\log_2w  + \log_2{k})\right)} \geq 
\]

\[
\geq 2^{kw((\log w)/6)-(1/(\log_2k\log_2w)^2)\log \left(C_1k(C_2 +0.5\log_2{w} - \log_2\log_2 k - \log_2\log_2w  + \log_2{k})\right)} >1 
\]

Therefore $SAF_{k,w} \not\in \kpobdd _{\sqrt{w}/ (\log_2 k \log_2w)}$, due to Lemma \ref{lm:lower-pnobdd}. 

And $\kpobdd _{\sqrt{w}/ (\log_2 k \log_2w)} \neq \kpobdd _{3w + 1}$ 
\Endproof

\section{Width Hierarchies for Quantum $k$-OBDD}\label{sec:s2}
Let us present existing lower bound for $k$-OBDDs.
\begin{lemma}[\cite{akk2017}]\label{lm:qobdd-lower-bound}
Let function $f(X)$ is computed by bounded error $k$-QOBDD P of width w, then $N^\pi(f) \leq w^{C\cdot(kw)^2}$ for some $C = const$.
\end{lemma}

Then let us discuss Boolean function {\em Matrix Xor Pointer Jumping}, which complexity property allow to show hierarchy.

\begin{definition}
Firstly, let us present version of $PJ$ function which works with integer numbers. Let $V_A, V_B$ be two disjoint sets (of vertices) with $|V_A| = |V_B| = d$ and $V = V_A \cup V_B$ . Let $F_A = \{f_A : V_A \to V_B\}$, $F_B = \{f_B : V_B \to V_A\}$ and $f = (f_A, f_B):V \to V$ defined by $f(v) = f_A(v)$, if $v\in V_A$ and $f= f_B(v)$, $v\in V_B$. For each $j \geq 0$ define $f^{(j)}(v)$ by $f^{(0)}(v) = v$ , $f^{(j+1)}(v) = f(f^{(j)}(v))$. Let $v_0\in V_A$. The functions we will be interested in computing is $g_{k,d} : F_A \times F_B \to V$ defined by $g_{k,d}(f_A, f_B) = f^{(k)}(v_0)$.

Definition of {\em Matrix XOR Pointer Jumping function} looks like {\em Pointer Jumping function}. 

Firstly, we introduce definition of $MatrixPJ_{2k,d}$ function. Let us consider functions $f_{A,1},f_{A,2}, \cdots f_{A,k}\in F_A$ and $f_{B,1},f_{B,2}, \cdots f_{B,k} \in F_B$.

%

On  iteration $j+1$ function $f^{(j+1)}(v)=f_{j+1}(f^{(j)}(v))$, where 

$f_i(v)= \begin{cases}
 & f_{A,\lceil \frac{i}{2} \rceil}(v)\text{ if } i \text{ is odd } \\ 
 & f_{B,\lceil \frac{i}{2} \rceil}(v)\text{ if } i \text{ is even }
\end{cases} $.

$MatrixPJ_{2k,d}(f_{A,1},f_{A,2}, \cdots f_{A,k}, f_{B,1},f_{B,2}, \cdots f_{B,k}) = f^{(k)}(v_0)$.

Secondly, we add $XOR$-part to $Matrix PJ_{2k,d}$ (note it $XMPJ_{2k,d}$). Here we take $f^{(j+1)}(v)=f_{j+1}(f^{(j)}(v)) \oplus f^{(j-1)}(v) $, for $j\geq 0$

Finally, we consider boolean version of these functions. Boolean function $PJ_{t,n}:\{0,1\}^n\to\{0,1\}$ is boolean version of $g_{k,d}$, where we encode $f_A$ in a binary string using $d\log d$ bits and do it with  $f_B$ as well. The result of function is parity of binary representation of result vertex.

In respect to boolean $MXPJ_{2k,d}$ function we encode functions in input in following order $f_{A,1}, \dots, f_{A,k}, f_{B,1}, \dots, f_{B,k}$. Let us describe process of computation on Figure \ref{fig:ab}. Function $f_{A,i}$ is encoded by $a_{i,1},\cdots a_{i,d}$, for $i \in \{1 \cdots k\}$. And $f_{B,i}$ is encoded by $b_{i,1}, \cdots b_{i,d}$, for $i \in \{1 \cdots k\}$. Typically, we will assume that $v_0=0$.

\begin{figure}[tbh]
\begin{center}
\includegraphics[width=5cm]{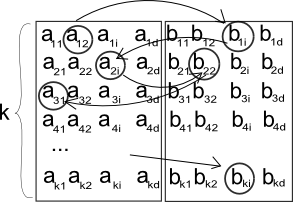}
\end{center}
\caption{Boolean function $XMPJ_{k,d}$}
\label{fig:ab}
\end{figure}

\end{definition}

Let us discuss complexity properties of $MXPJ_{2k,d}$.

\begin{lemma}[\cite{akk2017}]\label{lm:MXPJ-subfunc}
For $kd\log d = o(n)$ following is right:
$N^{id}(MXPJ_{2k,d})\geq  d^{\lfloor d/3-1 \rfloor (k-3)}$
\end{lemma}
\begin{lemma}[\cite{akk2017}]\label{lm:MXPJ-upper-bound}
There is exact $k$-id-QOBDD $P$ of width $d^2$ which computes  $MXPJ_{2k,d}$.
\end{lemma}
Let us present Lemma \ref{lm:MXPJ-subfunc} in more useful form:

\begin{corollary}\label{cr:MXPJ-lower}
For integer $k=k(n)$, $w=w(n)$, $kd\log d = o(n)$ and Boolean function $MXPJ_{2k,d}$, the following statement is
right: $N^{id}(MXPJ_{2k,d})\geq  d^{dk/16}$
\end{corollary}

\subsection{Hierarchy Results for Quantum Models}

Now we can prove hierarchy results for \kqobdds.
\begin{theorem}\label{th:hi-q}
We have $k$-id-{\bf QOBDD}$ _{\sqrt{d / C_1k}} \subsetneq k$-id-{\bf QOBDD}$ _{d^2}$ for some $C_1 = const$ .
\end{theorem}
\Beginproof
It is clear that $k$-id-{\bf QOBDD}$ _{\sqrt{d / C_1k}} \subseteq k$-id-{\bf QOBDD}$ _{d^2}$. Let us proof inequality of these classes.

Due to Lemma \ref{lm:MXPJ-upper-bound}, $MXPJ_{2k,d} \in k$-id-{\bf QOBDD}$_{d^2}$. Let us show that $MXPJ_{2k,d} \notin k$-id-{\bf QOBDD}$_{\sqrt{d / Ck}}$. 

\[
\frac{N^{id}(MXPJ_{2k,d})}{\sqrt{d / C_1k}^{C(k\sqrt{d / C_1k})^2}} \geq 
\]
\[
\geq \frac{d^{dk/16}}{\sqrt{d / C_1k}^{C(k\sqrt{d / C_1k})^2}} = 
\]

\[
= 2^{\frac{dk}{16}\log d - C(k\sqrt{d / C_1k})^2 \log{\sqrt{d / C_1k}}} = 
\]
\[
= 2^{\frac{k}{16} \left( d\log d - \frac{16Ckd}{2C_1k}\log{d / C_1k} \right)}
\]
Let $C_1 = 8C$ for C from Lemma \ref{lm:qobdd-lower-bound}, then we have
\[
\frac{N^{id}(MXPJ_{2k,d})}{\sqrt{d / C_1k}^{C(k\sqrt{d / C_1k})^2}} \geq 2^{\frac{k}{16} \log{C_1k}} > 1.
\]
Therefore $MXPJ_{2k,d} \notin k$-id-{\bf QOBDD}$_{\sqrt{d / Ck}}$, due to Lemma \ref{lm:qobdd-lower-bound}. 

And $k$-id-{\bf QOBDD}$ _{\sqrt{d / C_1k}} \neq k$-id-{\bf QOBDD}$ _{d^2}$.
\Endproof

\section{Discussion on Relation Between Models}\label{sec:s3}
You can see some existing discussion between models in \cite{akk2017}, \cite{agky16}, \cite{agkmp2005}, \cite{g15}, \cite{gy2017}, \cite{iky2017}. Here we will present some relations on fixed $k$.

Relation between models follows from Theorems \ref{th:hi-n}, \ref{th:hi-p}:
\begin{theorem}
There are Boolean function $f$, such that:

$f\in k$-{\bf OBDD}$_{3w + 1}, \knobdd _{3w + 1}, \kpobdd _{3w + 1};$

$f\not\in k$-{\bf OBDD}$_{\lfloor w/16
\rfloor-3}, \knobdd _{\sqrt{w} / 2},  \kpobdd _{\sqrt{w}/ (\log_2 k \log_2w)}$
\end{theorem} 

Let us compare classical models and quantum models.
Firstly, let us discuss classical complexity properties of $MXPJ_{2k,d}$ function.

\begin{lemma}
There is $k$-id-OBDD $P$ of width $d^2$ which computes  $MXPJ_{2k,d}$.
\end{lemma}
\Beginproof
Let us construct such $k$-id-OBDD $P$. 

By the definition of function $MXPJ_{2k,d}$ input separated into $2dk$ blocks by $t=\lceil \log_2 {d} \rceil$ bits. Blocks encode integers $a_{i1},a_{i2} \cdots a_{id}$ for $i \in \{1,\cdots k\}$ in the first part of input; and $b_{i1},b_{i2} \cdots b_{id}$ for $i \in \{1,\cdots k\}$ in the second part (see Figure \ref{fig:ab}). Let elements of block representing $a_{ij}$ be $X^{i,j}=(x^{i,j}_0, \dots, x^{i,j}_{t-1})$ for $i \in \{1,\cdots k\}, j \in \{1,\cdots d\}$ and elements of block representing $b_{ij}$ be $Y^{i,j}=(y^{i,j}_0, \dots, y^{i,j}_{t-1})$ for $i \in \{1,\cdots k\}, j \in \{1,\cdots d\}$

Let us discuss $i$-th layer. On the first level we have $d^2$ nodes, each of them corresponds to pair $(u,v)$, for $u,v\in {0,\dots, d-1}$ for storing $f^{(2i-3)}$ and $f^{(2i-2)}$. At first $P$ skips all blocks except $x^{i,f^{(2i-2)}}$. Then it will compute XOR of bits of the block and $u$ of pair. In the end of the block $P$ leads node corresponding to $(f^{(2i-1)}, f^{(2i-2)})$. After that the program skip all other blocks of first part and all blocks of first part except $y^{i,f^{(2i-1)}}$. Then it computes XOR with of bits of the block and $v$ of pair. In the end of the block $P$ leads node corresponding to $(f^{(2i-1)}, f^{(2i)})$.

On the last layer after computing $f^{(2k)}$, all nodes which XOR result is $1$  leads $1$-sink.
\Endproof

\begin{lemma}
$MXPJ_{2k,d}\not \in k$-id-{\bf OBDD}$_{d/32}$.
\end{lemma}
\Beginproof
Let us apply Lemma \ref{lm:lower-kobdd}
\[
	\frac{MXPJ_{2k,d}}{(d/32)^{(k-1)d/32+1}} 
	\geq \frac{d^{dk/16}}{(d/r)^{(k-1)d/32+1}}\geq
\]
\[
\geq 2^{\frac{kd}{16}\log d - 2\log(d/32)kd/(32)} 
= 2^{kd(\frac{\log d }{16}- 2(\log d- 5)/32)} >1 
\]
Therefore $MXPJ_{2k,d} \not\in k$-id-{\bf OBDD}$_{w/32}$, due to Lemma \ref{lm:lower-kobdd}.
\Endproof

\begin{lemma}
$MXPJ_{2k,d}\not \in k$-id-{\bf NOBDD}$_{\sqrt{(d\log d)/33}}$.
\end{lemma}
\Beginproof
Let us apply Lemma \ref{lm:lower-knobdd}. Let $r=\sqrt{33d/\log d}$
\[
	\frac{MXPJ_{2k,d}}{2^{((k-1)d/r+1)d/r}} 
	\geq \frac{d^{dk/16}}{2^{((k-1)d/r+1)d/r}}\geq
\]
\[
\geq 2^{\frac{kd}{16}\log d - 2(kd/r)d/r} 
= 2^{kd(\frac{\log d }{16}- 2d/r^2)} >1 
\]
Therefore $MXPJ_{2k,d} \not\in k$-id-{\bf NOBDD}$_{\sqrt{(d\log d)/33}}$, due to Lemma \ref{lm:lower-knobdd}.
\Endproof

\begin{lemma}
$MXPJ_{2k,d}\not \in k$-id-{\bf POBDD}$_{\sqrt{d/\log k}}$.
\end{lemma}
\Beginproof
Let us apply Lemma \ref{lm:lower-pnobdd}. Let $r=\sqrt{d/\log k}$
\[
	\frac{MXPJ_{2k,d}}{\left(C_1k(C_2 +\log_2{d} -\log_2{r}+ \log_2{k})\right)^{(k+1)d^2/r^2}} \geq
\]
\[
	\geq \frac{d^{dk/16}}{\left(C_1k(C_2 +\log_2{d} -\log_2{r}+ \log_2{k})\right)^{(k+1)d^2/r^2}}\geq
\]
\[
\geq 2^{\frac{kd}{16}\log d - 2(kd^2/r^2)\left(C_3 +\log_2 k + \log_2(C_2 +\log_2{d} + \log_2{k})\right)}=
\]
\[
= 2^{2kd(\frac{\log d }{32}- d\left(C_3 +\log_2 k + \log_2(C_2 +\log_2{d} + \log_2{k})\right)/r^2)} >1 
\]
Therefore $MXPJ_{2k,d} \not\in k$-id-{\bf POBDD}$_{d/\log k}$, due to Lemma \ref{lm:lower-pnobdd}.
\Endproof

Then base on these lemmas we can get following result:

\begin{theorem}
There is Boolean function $f$, such that:

$f\in k$-id-{\bf OBDD}$_{d^2}, k$-id-{\bf NOBDD}$_{d^2}$, $k$-id-{\bf POBDD}$_{d^2}, k$-id-{\bf QOBDD}$_{d^2};$

$f\not\in k$-{\bf OBDD}$_{\lfloor d/32
\rfloor-3}$, $ k$-id-{\bf NOBDD}$_{\sqrt{(d\log d)/33}}$, $k$-id-{\bf POBDD}$_{\sqrt{d/\log k}}$,  $k$-id-{\bf QOBDD}$_{\sqrt{d / C_1k}}$
\end{theorem} 

\bibliographystyle{alpha}
\bibliography{tcs}


\end{document}